\documentclass[letter]{aa} % for the letters 
\usepackage{graphicx}
\usepackage{natbib}
\usepackage[breaklinks, colorlinks, citecolor=blue]{hyperref}

\usepackage{txfonts}
\begin{document}

   \title{Discovery of a Thirty-Degree Long Ultraviolet Arc in Ursa Major}

   \author{A. Bracco\inst{*1,2}, R. A. Benjamin\inst{3}, M.I.R. Alves\inst{4}, A. Lehmann\inst{2}, F. Boulanger\inst{2}, L. Montier\inst{5}, D. Mittelman\inst{6}, D. di Cicco\inst{6}, \and S. Walker\inst{6}
          }
  \institute{Rudjer Bošković Institute, Bijenička cesta 54, 10000 Zagreb, Croatia \\
                \email{*abracco@irb.hr}
                \and
  Laboratoire de Physique de l’Ecole Normale Supérieure, ENS, Université PSL, CNRS, Sorbonne Université, Université de Paris, Paris, France
         \and
              University of Wisconsin-Whitewater, Department of Physics, 800 West Main St, Whitewater, WI, 53190, USA 
\and 
Radboud University, Department of Astrophysics/IMAPP, P.O. Box 9010, 6500 GL Nijmegen, The Netherlands 
\and 
IRAP, Université de Toulouse, CNRS, CNES, UPS, Toulouse, France
\and
MDW Sky Survey, New Mexico Skies Observatory, Mayhill, NM 88339, USA \\
             %\thanks{The university of heaven temporarily does not
             %        accept e-mails}
             }

   \date{Received: March 17, 2020; Accepted: April 3, 2020}

  \abstract
   {Our view of the interstellar medium of the Milky Way and the universe beyond is affected by the structure of the local environment in the Solar neighborhood. Here, we present the discovery of a thirty-degree long arc of ultraviolet emission with a thickness of only a few arcminutes: the Ursa Major Arc. It consists of several arclets seen in the near- and far-ultraviolet bands of the GALEX satellite. A two-degree section of the arc was first detected in the H$\alpha$ optical spectral line in 1997; additional sections were seen in the optical by the team of amateur astronomers included in this work. This direction of the sky is known for very low hydrogen column density and dust extinction; many deep fields for extra-galactic and cosmological investigations lie in this direction. Diffuse ultraviolet and optical interstellar emission are often attributed to scattering of light by interstellar dust. The lack of correlation between the Ursa Major Arc and thermal dust emission observed with the {\it Planck} satellite, however, suggests that other emission mechanisms must be at play. We discuss the origin of the Ursa Major Arc as the result of an interstellar shock in the Solar neighborhood.}

\authorrunning{A. Bracco et al.}
\titlerunning{Discovery of a Thirty-Degree Long Ultraviolet Arc in Ursa Major}    
\maketitle

%
%-------------------------------------------------------------------
\section{Introduction}\label{sec:intro}
New observations of the diffuse interstellar medium (ISM) in 21-cm emission, dust emission and polarization, and low radio frequency polarized emission, have awakened new interest in linear structures in the ISM that may affect our view of the Solar neighborhood \citep{Clark2015,Jelic2015J,PlanckXXXII}. The study of the structure of the local and diffuse ISM is important as it provides a foreground that must be accounted for in studying our Milky Way and the universe beyond.

In 2001 P. McCullough \& R.A. Benjamin presented the discovery of an unusually straight and narrow ionized $2^{\circ}.5$-long filament detected in H$\alpha$ (656 nm, hereafter the MB filament) — with a surface brightness of $\sim$0.5 R — in the direction of Ursa Major \citep[][hereafter MB01]{MB2001}, whose origin is yet to be established. They considered four options for the origin of this structure: (1) a low-density jet, (2) an unusually straight nebular filament, (3) shock ionization by a compact source, e.g. a neutron star, and (4) a trail of ionized gas left by a low-luminosity ionizing source. Although they favored the final "Fossil Stromgren Trail" (FST) option, no plausible ionizing source was ever identified.

In this letter we present the outstanding discovery of a longer portion of the MB filament seen both in the ultraviolet and in H$\alpha$ extending over thirty degrees on the sky. We discuss its origin in terms of interstellar shocks in the ISM.   

%--------------------------------------------------------------------
\begin{figure*}[!ht]
\centering
\includegraphics[width=1.\textwidth]{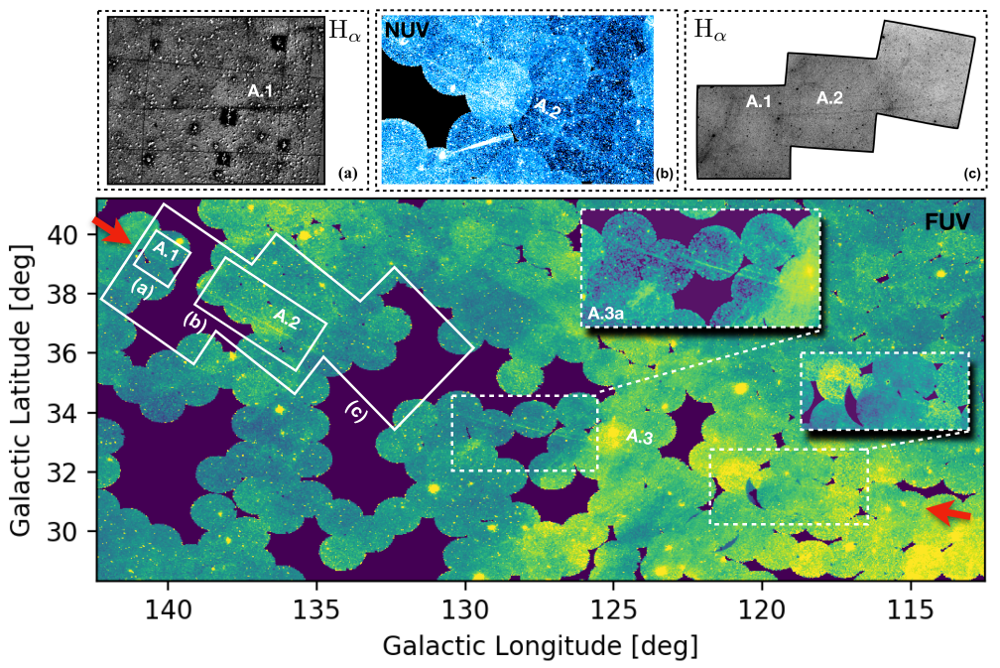}
\caption{{\it The Ursa Major (UMa) Arc, Feature A (between the red arrows), in a mosaicked image of GALEX FUV data taken in 2005. Each (circular) pointing has a diameter of $1^{\circ}.27$.  Inset (a) shows the original 1997 detection of the H$\alpha$ filament published by \cite{MB2001}; inset (b) shows part of the arc in GALEX NUV data; inset (c) shows a section of the arc that was observed in H$\alpha$ as part of the MDW Hydrogen-Alpha Sky Survey in 2016 \citep{MDW}. The bright NUV feature below A.2 is an artifact in the GALEX data. Intensity scaling in all diagrams is arbitrary. In general, brighter colours correspond to more intense regions in the FUV and NUV maps, while darker colours correspond to more intense H$\alpha$. We label the three sections of this feature as A.1, A.2, and A.3 interrupted by gaps in the GALEX coverage. The two additional insets in the FUV maps along A.3 have an adapted color scale to optimize the visualization of the arcs. 
}}
\label{fig:UMa}
\end{figure*}

\section{Description of the Ursa Major (UMa) Arc}\label{sec:UMa}

We investigated the MB filament in several wavebands using the Aladin sky-viewer \citep{Boch2014} and discovered that the MB filament is also visible both in the Far Ultraviolet (FUV; 130-180 nm) and in the Near Ultraviolet (NUV; 170-280 nm) bands of the Galaxy Evolution Explorer (GALEX) All-Sky Imaging Survey (AIS) \citep{Bianchi2009}. The Ultraviolet imaging, which covers a much wider area of the sky than the original H$\alpha$ observations, demonstrates that the MB filament is a piece of a much larger arc-like structure. A GALEX FUV image displaying a large fraction of the Ursa Major Arc (UMa Arc) is shown in Fig.~\ref{fig:UMa} along with zoom-in views of selected sections in H$\alpha$ and the GALEX NUV band. The structure consists of several piecewise-continuous arclets, occasionally multiple. Most of the arclets we found lie along a curved path from Galactic coordinates ($l,b$) = (115$^{\circ}$.4, 30$^{\circ}$.7) to (147$^{\circ}$.7, 43$^{\circ}$.5).

If the structure were due to a linear trail, the option favored by MB01, it would project onto a great circle on the sky. We found that no single great circle can fit the arc we observed. In addition, a search of GALEX images of nearby white dwarfs and hot subdwarfs with parallaxes and proper motions measured by the Gaia satellite \citep{Geier2019} shows no structures in the FUV similar to the original MB filament. For these two reasons, the FST hypothesis appears to be ruled out. 

We have found instead that most of the arclets lie along one-sixth (1 radian) of a circle centered on ($l,b$) = (107$^{\circ}$.7, 60$^{\circ}$.0) with angular radius $\Theta_{\rm R} = 29^{\circ}.28 \pm 0^{\circ}.23$, labeled as Feature A in Fig.~\ref{fig:UMa}. If one three-degree long arclet, centered at ($l,b$) = (131$^{\circ}$,+33$^{\circ}$.7) (the lowest arclet in A.3a in Fig.~\ref{fig:UMa}), is not considered, the standard deviation drops to $\sigma_{\Theta} = 0^{\circ}.18$ or 0.6\% of the angular radius. The individual arclets are even thinner, ranging from 0.3 to a few arcminutes. An extrapolation of this arc — the UMa Circle — encircles a total area of 2681 square degrees (0.25$\pi$ steradians), 6.5\% of the total sky and 27\% of the sky above Galactic latitude b = 32$^{\circ}$. The angular size of this full region is comparable to the largest angular size radio-continuum loops discovered in the 1960s \citep{Quigley1965}. What distinguishes the UMa Arc compared to these other structures is (1) the thickness-to-radius ratio, (2) the nearly perfect circular fit over thirty degrees of arc, and (3) the high Galactic latitude of the center. 

\begin{figure*}[!ht]
\centering
\includegraphics[width=1.\textwidth]{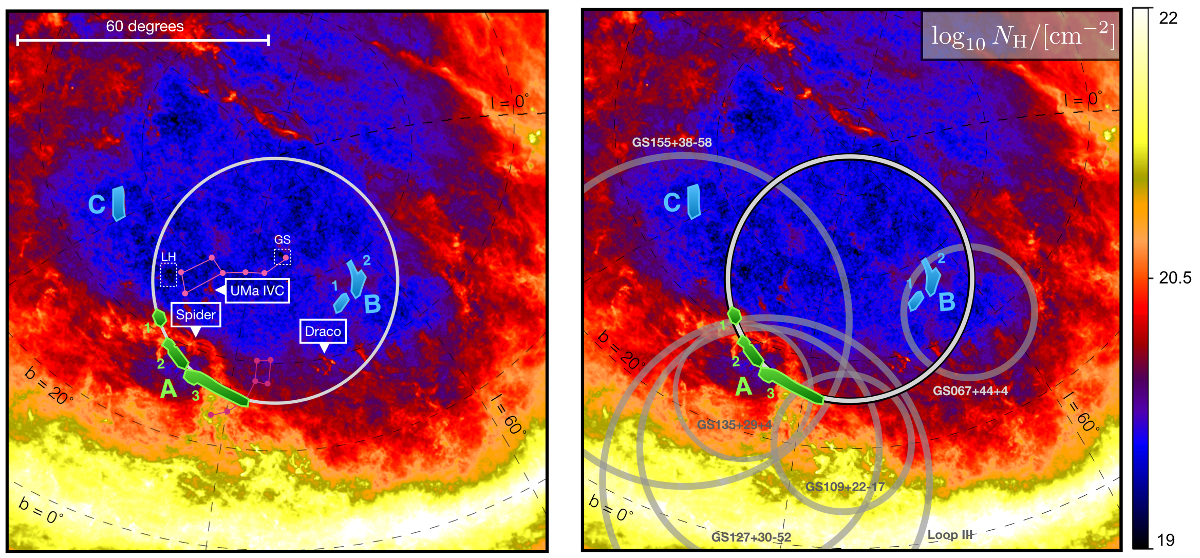}
\caption{{\it Stereographic projection of the total gas column density, $N_{\rm H}$, derived from Planck observations of dust thermal emission (color scale). The locations of Feature A, which defines the UMa Arc, and  Features B and C are marked. The UMa Circle is shown in light gray. The left panel gives the positions of the UMa IVC, Draco, and Spider clouds, along with the Ursa Major and Minor constellations (in pink). Two of the lowest column-density windows in the Galaxy (Lockman hole (LH) and Groth strip (GS)) are shown. The right panel indicates the positions of known arches and loops, in dark gray. The transition between blue and red on the false-colour intensity scale corresponds to $N_{\rm H} = 10^{20}$\,{\rm cm}$^{-2}$.}}
\label{fig:stereo}
\end{figure*}

Figure~\ref{fig:stereo} shows the positions of the arclets and the UMa Circle on a stereographic projection, overlaid on the total gas column density derived from {\it Planck} observations of thermal dust emission \citep{P2013XI2014}. A visual search of the GALEX mosaics identified three additional high-latitude arclets: two (Feature B.1 and B.2, centered at ($l,b$) = (79$^{\circ}$,+52$^{\circ}$)) lie interior to the UMa circle; only one (Feature C, centered at ($l,b$) = (185$^{\circ}$,+53$^{\circ}$)) lies outside this region (see also Fig.~\ref{fig:BC}). Also overlaid in Fig.~\ref{fig:stereo} are the locations of several previously claimed loops, shells, and arcs based on radio continuum observations (Radio Loop III) \citep{Quigley1965}, studies of dust emission (North Celestial Loop/GS135+29+4) \citep{Meyerdierks1991}, and 21-cm observations (Intermediate Velocity Arch/GS155+38-58 \citep{Kuntz1996} plus additional Heiles shells \citep{Heiles1984}). None of these shells — which are far less well delineated than the UMa Arc — are clearly aligned with the arc we observe. We also investigated the Effelsberg-Bonn HI Survey (EBHIS) \citep{Winkel2016} and found no evidence for a 21-cm counterpart to the H$\alpha$, or ultraviolet, emission. For visualization purposes we also present the UMa Arc and Circle in Fig.~\ref{fig:overview}.

The region within the UMa Circle is well known for having the lowest column densities of neutral hydrogen in the sky, including the "Lockman Hole" \citep{Lockman1986}, which lies 5$^{\circ}$.4 interior to the arc. This low column density region is part of a vertically extended, tilted low-density cavity identified as the "Local Chimney" \citep{Welsh1999}. The low column density and the large angular size of the UMa Circle also make it a potentially important foreground for extra-galactic and cosmological studies, requiring a thorough characterization of the interstellar emission interior and exterior to the arc. 

Locating the UMa Arc along the line of sight is not trivial given the data in hand. The large angular size suggests its proximity. The presence of three well-known interstellar clouds that lie — in projection — within the UMa Circle may be also indicative: the UMa molecular intermediate-velocity cloud (IVC), the Spider cloud, and the Draco molecular IVC, with distances of 356$\pm$20 pc, 369$\pm$22 pc, and 481$\pm$50 pc, respectively \citep{Zucker2019}. If the Draco cloud were physically inside the UMa Arc, it would imply that the center of this structure is more than 350 pc above the Galactic midplane with a radius larger than 200 pc. Given all these caveats and the large angular size of the structure, the UMa Arc cannot be further than a few hundred pc from the Sun.
%--------------------------------------------------------------------
\section{The ultraviolet emission of the UMa Arc}\label{sec:UV}

\begin{figure}[!ht]
\centering
\includegraphics[width=0.5\textwidth]{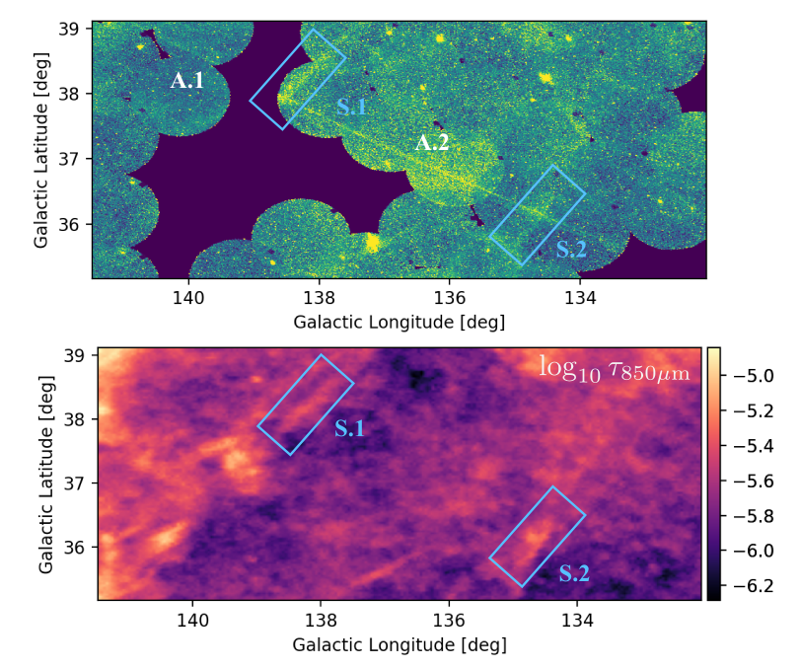}
\caption{{\it Comparison between the GALEX FUV data (top panel) and the dust optical depth ($\tau_{850 \mu{\rm m }}$, bottom panel), derived from the dust thermal emission measured by Planck at 850\,$\mu$m. Regions S.1 and S.2 are identified in both FUV and dust emission, whereas features A.1 and A.2 are only seen in FUV emission.}}
\label{fig:100micron}
\end{figure}

In the following, we examine the source of the UV emission from the UMa Arc. Analyses of high latitude diffuse ultraviolet emission using both FIMS/SPEAR (on the STSAT-1 korean satellite) and GALEX have shown that FUV emission is generally highly correlated with 100 $\mu$m dust emission — due to dust scattering — and somewhat correlated with H$\alpha$ intensity \citep{Seon2011}. Figure~\ref{fig:100micron} shows the same section of Feature A in FUV (top panel) and in dust optical depth (bottom panel) as traced by the {\it Planck} observations at 850 $\mu$m. While areas of enhanced dust emission, S.1 and S.2, in the {\it Planck} image are associated with an increase in FUV brightness, presumably due to dust scattering, the UMa Arc features, A.1 and A.2, are clearly not associated with dust emission. This is also confirmed by the average FUV/NUV ratio that we computed along the A.2 portion of the Arc. As detailed in Appendix \ref{app:ratio}, we measured the ratio using perpendicular profiles to the FUV and NUV filaments in the GALEX maps and averaging them along their full length. We found a ratio of 0.31 $\pm$ 0.05, which is significantly smaller than the values expected for dust scattering, mostly larger than unity \citep{Murthy2010a}. 

If not associated with dust scattering, other sources of diffuse FUV emission must be considered including the two-photon continuum emission from hydrogen, ion-atomic emission lines, e.g., C IV $\lambda$ 154.8/155.1 nm, He II $\lambda$ 164.0 nm, and O III] $\lambda$ 166.2 nm, and fluorescent molecular hydrogen emission lines.  
%--------------------------------------------------------------------
\section{Is the UMa Arc a shock front?}\label{sec:shock}

\begin{figure*}[ht]
\centering
\includegraphics[width=1.\textwidth]{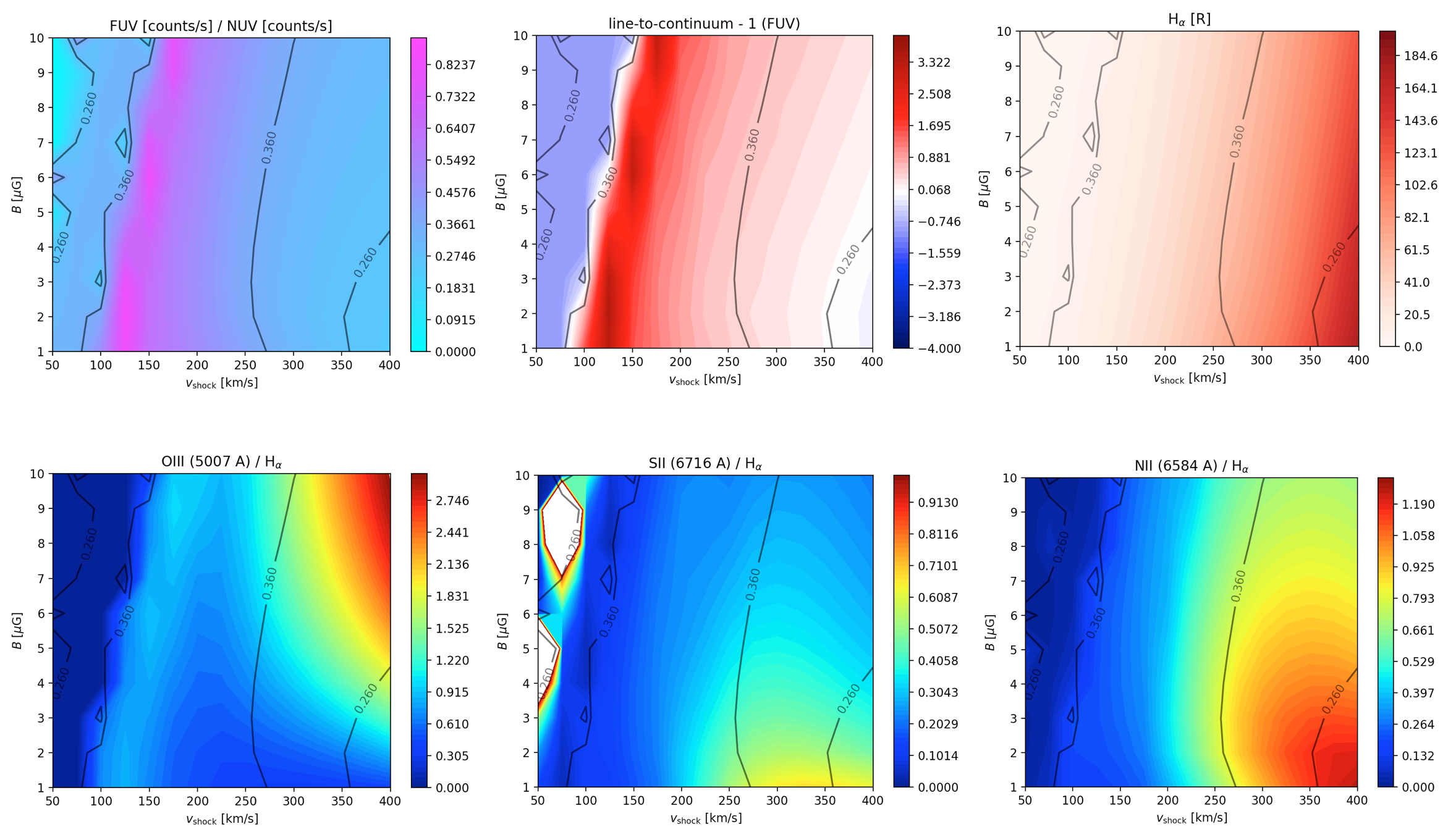}
\caption{\it Grid of models of radiative shocks from \cite{Sutherland2017} with varying magnetic-field strengths and shock velocities in a diffuse medium with pre-shocked density of $0.1$ cm$^{-3}$. Top row: FUV/NUV ratio from the models (left), line/continuum ratio in the GALEX FUV bandpass (center, similar behaviours are found in the NUV case), H$\alpha$ emission (right). The line to continuum ratio is centered around 0 so that blue (red) colours correspond to larger (smaller) continuum emission compared to line emission. Bottom row: ion line to H$\alpha$ ratios for OIII (left), SII (center), and NII (right). In all six panels the black contours represent the 1-$\sigma$ level of the observed FUV/NUV of $0.31 \pm 0.05$.}
\label{fig:models}
\end{figure*}

Ion lines are proxies of hot gas (with temperatures > $10^4$ K) likely produced in the cooling zone of interstellar shocks with velocities larger than 50 km/s \citep{Nishikida2006}. Recombination of the post-shock gas would produce as well two-photon emission contributing to the UV continuum. Radiative shocks are expected to be a key source of dissipation of kinetic energy driven by supernova blast waves in the ISM, making supernova remnants (SNRs) the most favorable scenario for the formation of structures such as the UMa Arc. 

To test this idea we inspected GALEX FUV images of four large angular-diameter ($\Theta_{\rm R} > 10^{\circ}$) old SNRs: the Monogem Ring \citep{Plucinsky1996}, Antlia SNR \citep{MB2002}, G70.0-21.5 \citep{Fesen2015} and G353-34 \citep{Testori2008}. The first two objects have been the targets of low angular resolution FUV spectroscopy with FIMS/SPEAR \citep{Shinn2007}. All four objects show networks of thin FUV filaments reminiscent of the UMa Arc, but over a smaller solid angle of the sky; two of these, Antlia SNR and G070.0-21.5, have also been shown to have similar H$\alpha$ filaments. In Fig.~\ref{fig:SNR} we show the case of G070.0-21.5 in  H$\alpha$, FUV, and dust emission. 

The geometry of the arclets, the emission in the GALEX NUV and FUV channels, and the concomitant H$\alpha$ are all consistent with the hypothesis that the UMa Arc is a shock, or pressure front, created by an explosion, yet unknown, towards ($l,b$) = (107$^{\circ}$.7,60$^{\circ}$.0). The thinness and multiplicity of the structures could be produced by the edge-on projection of corrugated sheets in the front and post-shock recombination zones, similar to what is seen in the Cygnus Loop SNR \citep{Hummer1987,Raymond1988}. If this hypothesis was true then the UMa Arc could be associated to a corrugated edge on sheet of the Loop III SNR (see Fig. ~\ref{fig:stereo}).  

With the caveat that the radius of curvature of an individual arc is not always the same as the radius of curvature of the full remnant, we can compare the UMa Arc to expectations of a simple model of blast wave evolution for a SNR \citep{Draine2011}. Under the assumption that the UMa Arc is a section of a front centered on a point $d_{100} = d/(100 {\rm pc})$ from the Sun and using a standard explosion energy of $E_{\rm SN} = 10^{51}$ ergs, the remnant will still be in the adiabatic phase — with shock speed greater than 190 km/s — if the ambient density is $n_{\rm H,0} < (0.17 {\rm cm}^{-3} ) d_{100}^{-2.38}$. Alternately, for an assumed ambient density of $n_{{\rm H},0} = 1$ cm$^{-3}$, the remnant would have transitioned to a pressure-modified momentum-conserving snowplow\footnote{In this phase the energy radiated behind the shock front is comparable to the initial energy of the explosion.} phase with an age of $t_{\rm snow} = (0.7\,{\rm Myr})d_{100}^{3.5}$  and a front speed that is approaching the random ISM velocity. 

Assuming that the ambient density was sufficiently high to trace a radiative front, we computed a series of shock models from publicly available libraries \citep{Sutherland2017} with shock velocities between 50 km/s and 400 km/s, pre-shock interstellar magnetic-field strengths between 1 $\mu$G and 10 $\mu$G, and mean diffuse gas density in the pre-shock medium of 0.1 cm$^{-3}$. These models allow us to predict the level of emission — both lines and continuum — within the GALEX bandpasses at the shock front that we can compare to the observed FUV/NUV ratio. 

As shown in the top row of Fig.~\ref{fig:models}, we identified two possible domains that reproduce the observations delimited by the 1-$\sigma$ contour of the measured FUV/NUV ratio. Shocks at intermediate velocities (between 50 km/s and 100 km/s) and at high velocity (> 300 km/s) are both consistent with the measured value of 0.31 $\pm$ 0.05. These two regimes, however, show opposite line-to-continuum ratios, that is, only at intermediate velocities (<100 km/s) the two-photon continuum emission dominates over the ion-line contribution. The level of two-photon emission predicted by these models is also consistent with the analytical derivation of the FUV/NUV count ratio of 0.37 that can be found in Appendix~\ref{app:2photon}. This intermediate velocity range is favored as well by the predicted H$\alpha$ surface brightness (see the top-right panel in Fig.~\ref{fig:models}) that compares well with the value of 0.5 R measured in 1997 \citep{MB2001}. Nonetheless, notice that our models do not take into account limb-brightening projection factors that alter the estimate of the H$\alpha$ surface brightness \citep{Hummer1987}. Additional constraints are therefore needed and may come from future spectral observations of ion lines (i.e. O III], SII, and NII) toward the UMa Arc, which, as predicted by our models and presented in the bottom row of Fig.~\ref{fig:models}, will reveal opposite line ratios relative to H$\alpha$ in the two velocity regimes. 

%--------------------------------------------------------------------
\section{Conclusions}\label{sec:conclusions}

In this letter we reported the stunning discovery of the UMa Arc, or one of the largest and diffuse structures of the Northern sky, yet unknown until now. We discussed the possibility that the UMa Arc is the relic of a radiative shock in the neighborhood of the Sun (< few hundred pc). Future work will be crucial to better constrain the physical parameters of the UMa Arc, as well as its distance, both for understanding our local astrophysical environment and for properly describing the structure of the Galactic ISM as a foreground for extra-galactic and cosmological studies. It will be necessary to (1) measure the H$\alpha$ radial velocity along the length of the Arc, (2) to constrain the proper motion of the MB filament since the original 1997 observations, and (3) search for associated interstellar absorption lines to bracket the distance to the structure. Finally, future observational data, including optical/UV emission line spectra to obtain shock parameters, will put additional constraints on the presumed front.

\begin{acknowledgements}
     We thank the anonymous referee for her/his comments. We would like to acknowledge Peter McCullough as the original discoverer of the MB filament which led to this work nearly twenty years later. We also thank the organizers of the 2018 program Milky-Way-Gaia of the PSI2 project funded by the IDEX Paris-Saclay, ANR-11-IDEX-0003-02, where this work started. We thank David Schiminovich for advice on calculating GALEX count-rates and John Raymond and Jon Slavin for useful discussions about GALEX data and supernova remnants. We would also like to acknowledge the hard work and untimely passing of David Mittelman in 2017, who strongly contributed to the realization of the MDW survey. We thank as well Bruce Draine and Peter G. Martin for comments to the manuscript. A.B. acknowledges the support of the Agence Nationale de la Recherche (project BxB: ANR-17-CE31-0022) and funding from the European Union’s Horizon 2020 research and innovation program under the Marie Skłodowska-Curie Grant agreement No. 843008. A.B. thanks as well the game of {\it la marelle} \url{https://www.youtube.com/watch?v=8c4tsQvRNWY} to help breaking his quarantine walls. M.I.R.A acknowledges funding from Radboud University (Excellence Fellowship); R.A.B. acknowledges the support of NASA grant NNX17AJ27G. This research has made use of "Aladin sky atlas" developed at CDS, Strasbourg Observatory, France. 
\end{acknowledgements}

\bibliographystyle{aa}
\bibliography{aanda.bbl}
\appendix
\section{Measuring the FUV/NUV ratio from the GALEX data}\label{app:ratio}
\begin{figure*}[!ht]
\centering
\includegraphics[width=1.\textwidth]{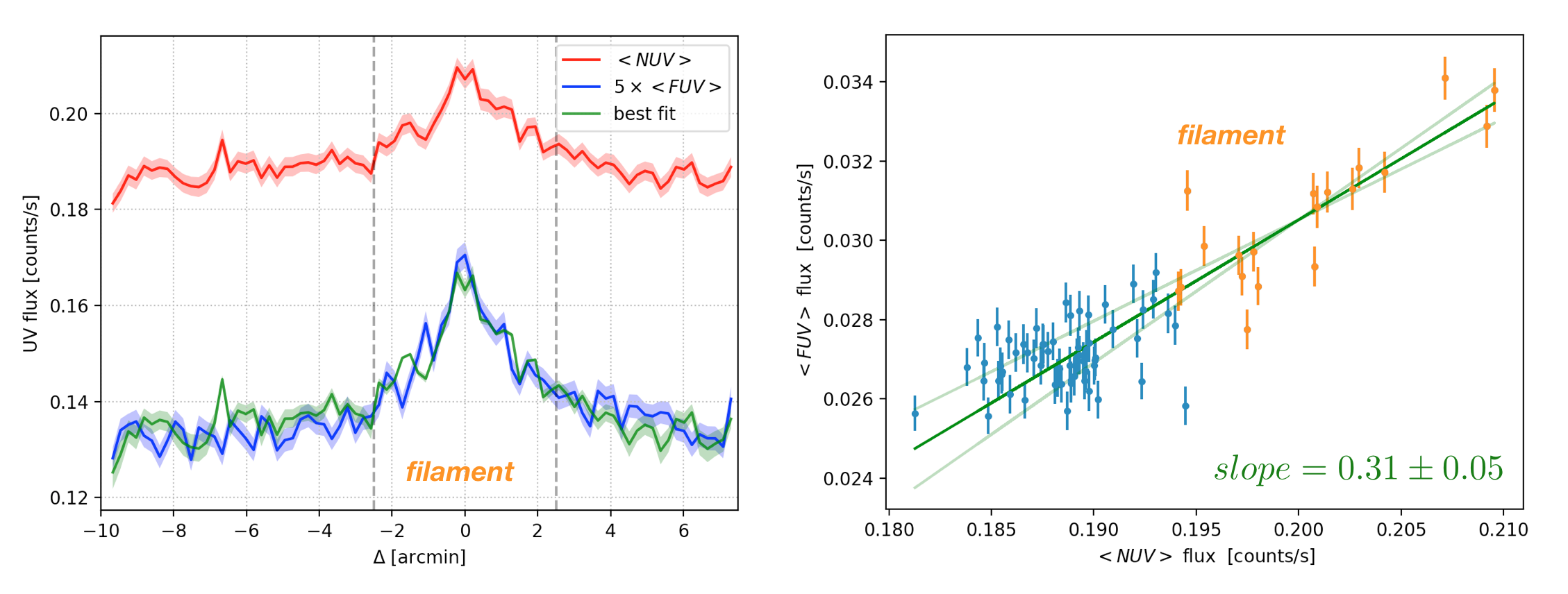}
\caption{\it Left panel: FUV (blue) and NUV (red) intensities, averaged over the whole length of A.2 (see Fig.~\ref{fig:100micron}), versus position across the width in arcmin. We show the average profiles, labeled $<FUV>$ and $<NUV>$, and their corresponding errors. $<FUV>$ is scaled up for visualization purposes, as well as the best-fit profile shown in green (see main text and right panel). Two vertical dashed lines delimit the borders of the A.2 filament merging with the local background/foreground UV emission. Right Panel: scatter plot and linear correlation between $<FUV>$ and $<NUV>$. The linear fit was computed only considering the data points (see orange circles) within the vertical dashed lines in the left panel. The intensity profile obtained scaling the $<NUV>$ with the best-fit parameters of the linear fit is shown in green on the left.}
\label{fig:corr}
\end{figure*}

We computed the ratio between the FUV and NUV bands for the straightest/longest portion of the UMa Arc (A.2 in Fig.~\ref{fig:UMa}) using the mosaics of the GALEX All-Sky Survey that were delivered by the GALEX team to the Mikulski Archive for Space Telescopes (MAST) and can be found at \url{https://archive.stsci.edu/prepds/gcat/gcat_dataproducts.html} with further documentation at \url{http://www.galex.caltech.edu/wiki/GCAT_Manual}. 

In order to account for visible offset variations among the different tessellations of the mosaic (see Fig.~\ref{fig:UMa}), we computed the FUV/NUV ratio by correlating average UV fluxes perpendicular to the main crest of A.2 (see Fig.~\ref{fig:corr}), labeled < FUV > and < NUV >, respectively. With the caveat that offsets were uniform within each tessellation, any miscalibration among the mosaics would be captured by the intercept of the linear fit between < NUV > and < FUV >, while the slope would correspond to the actual ratio between the two frequency bands. 

In the right panel of Fig.~\ref{fig:corr} we show the linear fit and the corresponding slope of 0.31 $\pm$ 0.05 obtained for the pixels that correspond to the position of A.2 (orange data points). As shown by the green profile in the left panel of the same figure, the linear fit allows us to strikingly reproduce the observed < FUV > profile from the observed < NUV >. 

\section{Theoretical estimate of the two-photon emission}\label{app:2photon}

Below we calculate the expected GALEX countrate for two-photon emission, which is expected to produce a fixed ratio of FUV to NUV emission \citep{Spitzer1951}. For two-photon emission, the Case B emissivity per unit frequency (${\rm ergs~s^{-1}~cm^{-2}~sr^{-1}~Hz^{-1}}$) is given by \citep{Draine2011}

\begin{equation}
    j_{\nu}^{B}(2s \rightarrow 1s)=j_{\nu}^{2s,B}=\frac{1}{4\pi}n_{e}n_{p}\alpha_{eff}^{(2s)B}(T)\left[h\nu P_{\nu}^{2s} \right] ,
\end{equation}

\noindent where $n_{e},n_{p}$ are the electron and proton densities, $\alpha_{eff}^{(2s)B}(T)$ is the total recombination to the hydrogen $2s$ level, where we use the data for Case B recombination \citep{Hummer1987}, and $[h\nu P_{\nu}^{2s}]$ is the energy emitted per frequency interval for the two-phton continuum. The recombination rate to the $2s$ level is one-third of the total recombination rate, where we use a fit to the temperature dependence \citep{Draine2011}: $\alpha_{eff}^{(2s)B}(T)=(1/3)(2.59 \times 10^{-13}~{\rm cm^{3}~s^{-1}})~f^{2s}(T)$ where $f^{2s}(T)=T^{-0.833-0.035\ln T_{4}}$ and $T_{4}=T/10^{4}$ K. We use an analytical fit to the spectral shape of the two photon continuum \citep{Nussbaumer1984}: $\left[h\nu P_{\nu}^{2s} \right]=hyA(y)/A_{2s\rightarrow 1s}$, where $y=(\nu/\nu_{Ly~\alpha})$ is the frequency scaled by the Lyman $\alpha$ frequency, $A_{2s\rightarrow 1s}=8.2249~{\rm s^{-1}}$ is the Einstein A value for the two-photon decay, and 
\begin{equation}
    A(y)=C \left[ y(1-y)(1-(4y(1-y))^{\gamma} +\alpha(y(1-y))^{\beta}(4y(1-y))^{\gamma} ) \right], 
\end{equation}

\noindent with $C=202.0~{\rm s^{-1}}$, $\alpha=0.88$, $\beta=1.53$, and $\gamma=0.8$. It is also convenient to define the unitless function $B(y)=A(y)/202.0~{\rm s^{-1}}$. 

Taken together, for a constant density path length, $L$, with an emission measure of $EM=n_{e}n_{p}L$, the two-photon continuum intensity per frequency interval, $I^{2s,B}_{\nu}=\int j^{2s,B}_{\nu}ds$ is

\begin{equation}
\begin{split}
   I^{2s,B}_{\nu}(y,T,EM)={\rm (3.45 \times 10^{-21}~ergs~s^{-1}~cm^{-2}~sr^{-1}~Hz^{-1})}\times \\ \times f^{2s}(T)yB(y) \\ 
   \left[ \frac{EM}{{\rm cm^{-6}~pc}} \right]
   \end{split}
\end{equation}

Similarly, the Case B recombination H$\alpha$ intensity can be written as 

\begin{equation}
\begin{split}
   I^{H\alpha,B}(T,EM)={\rm (8.71 \times 10^{-8}~ergs~s^{-1}~cm^{-2}~sr^{-1})}~f^{H\alpha}(T) \\ \left[ \frac{EM}{{\rm cm^{-6}~pc}} \right],
\end{split}
\end{equation}

\noindent where $f^{H\alpha}(T)=T_{4}^{-0.942-0.031\ln T_{4}}$. Expessed in Rayleighs (R), the H$\alpha$ intensity is $I^{H\alpha,B}(T,EM)=(0.36 ~R)f^{H\alpha}(T) ~[EM/{\rm cm^{-6}~pc}]$. For $T=8000$ K, this yields the oft-used conversion that $1 R$ corresponds to $EM=2.25~{\rm cm^{-6}~pc}$.

In wavelength units, $\lambda_{\text{\AA}}=\lambda/(1 \text{\AA})$, where $y=1215/\lambda_{\text{\AA}}$, the two-photon continuum intensity is given by 

\begin{equation}
\begin{split}
   I^{2s,B}_{\lambda}(y,T,EM)={\rm (1.04 \times 10^{-2}~ergs~s^{-1}~cm^{-2}~sr^{-1}~\text{\AA}^{-1})}\times \\
   \times f^{2s}(T)yB(y)\lambda_{\text{\AA}}^{-2} \left[ \frac{EM}{{\rm cm^{-6}~pc}} \right].
\end{split}
\end{equation}

To calculate the bandpass weighted continuum intensities, $I_{band}=\int T_{band}(\lambda)I^{2s,B}_{\lambda}d\lambda~/~\int T_{band}(\lambda)d\lambda$ we use the intensity formula above together with the effective area curves, $T_{band}(\lambda)$, from \url{http://www.galex.caltech.edu/researcher/effective_area/}. For an emission measure of $EM=1~{\rm cm^{-6}~pc}$ at $T=8000$ K, this yields $I_{band}=3.6$ and $1.4 \times 10^{-10}~{\rm ergs~s^{-1}~cm^{-2}~\text{\AA}^{-1}~sr^{-1}}$ for the FUV and NUV channels, respectively. Using the standard GALEX conversion \citep{Hamden2013} from energy units ${\rm (ergs ~s^{-1}~cm^{-2}~\text{\AA}^{-1})}$ to countrate ${\rm (counts~s^{-1})}$, and assuming a pixel size of $(12.88)^2~{\rm arcsec}^2$ yields a predicted GALEX countrate of $C_{band}=(0.98~{\rm and}~2.61 \times 10^{-3}~{\rm counts~s^{-1}~pixel^{-1}})~[EM/{\rm cm^{-6}~pc}]$ 
for the FUV and NUV channels, resulting in a predicted FUV/NUV count ratio for two-photon continuum emission of $0.37$. 

\section{Supporting figures}\label{app:overview}
In this appendix we show a series of figures to support the main body of the paper. 

\begin{figure}[ht]
\centering
\includegraphics[width=0.5\textwidth]{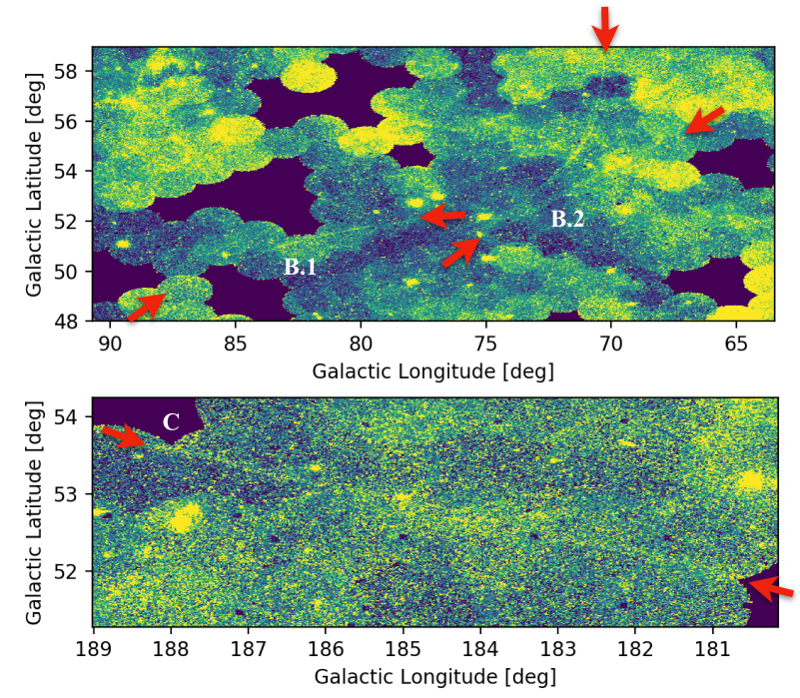}
\caption{\it Features B (top) and C (bottom) in a mosaicked image of GALEX FUV data.}
\label{fig:BC}
\end{figure}

Figure~\ref{fig:BC} shows Features B and C introduced in Sect.~\ref{sec:UMa}.  

\begin{figure*}[ht]
\centering
\includegraphics[width=0.7\textwidth]{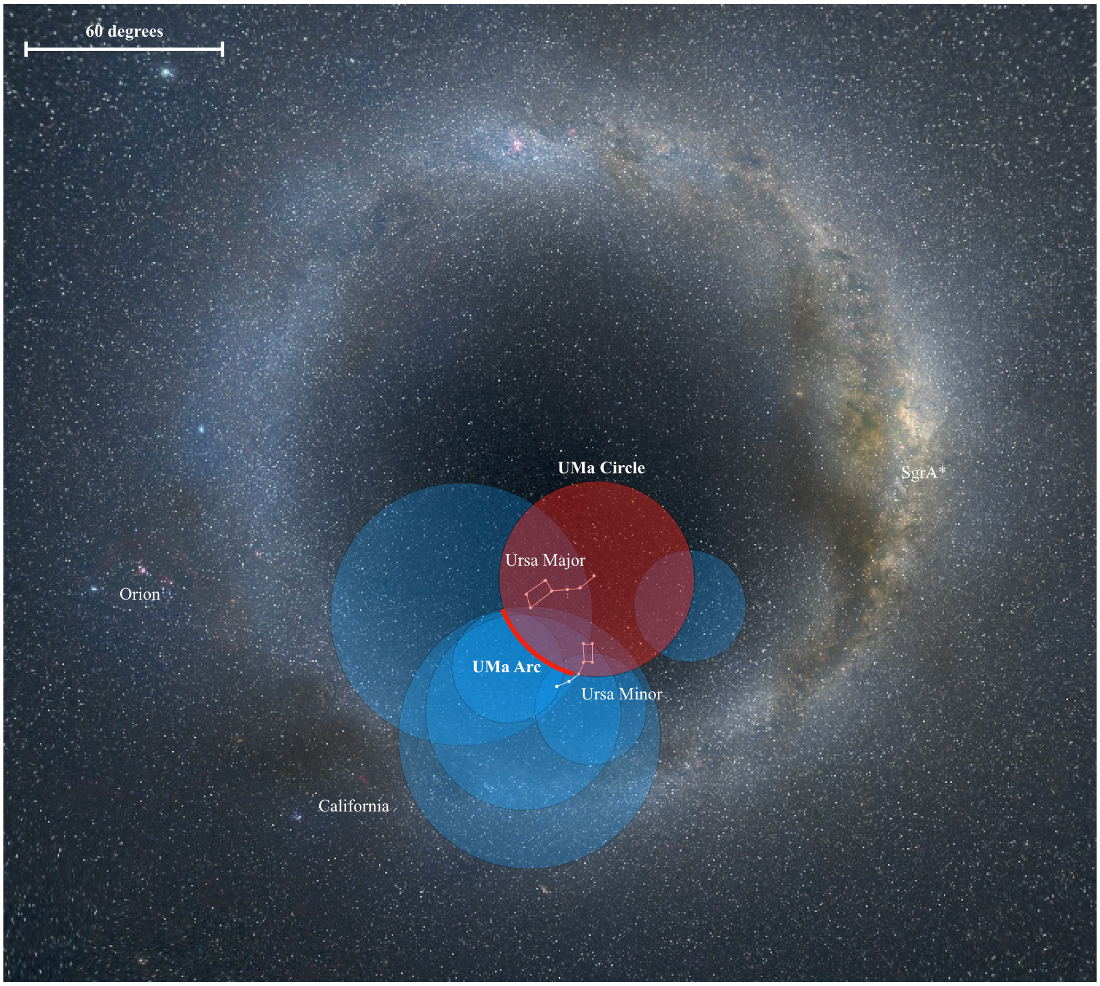}
\caption{\it Large-scale stereographic overview of the UMa Arc and Circle (red) on top of the Mellinger full-sky map. We also show as reference the positions of the loops and shells presented in Fig.~\ref{fig:stereo} in light blue as well as some known constellations and the Galactic center (SgrA*).}
\label{fig:overview}
\end{figure*}

In Fig.~\ref{fig:overview}, for visualization purposes, we show a large-scale stereographic projection of the UMa Arc and Circle (in red) on top of the Mellinger full-sky map \citep{Mellinger2009}. The loops and shells already presented in Fig.~\ref{fig:stereo} are shown in light blue. 

\begin{figure*}[ht]
\centering
\includegraphics[width=1.\textwidth]{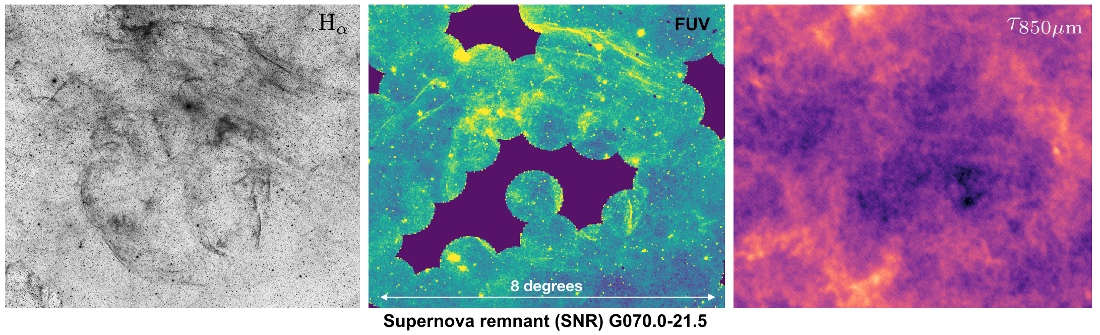}
\caption{\it SNR G070.0-21.5 (see Sect.~\ref{sec:shock}) in H$\alpha$ from the MDW survey (left), in FUV from GALEX (centre), and in dust optical depth from {\it Planck} (right).}
\label{fig:SNR}
\end{figure*}

For comparison with the UMa Arc, Fig.~\ref{fig:SNR} presents the SNR G070.0-21.5 introduced in Sect.~\ref{sec:shock}. It shows the SNR in H$\alpha$ from the MDW survey, in FUV from GALEX, and in dust optical depth from {\it Planck}. As for the UMa Arc, an intertwined network of filaments can be seen in H$\alpha$ and FUV without any counterpart in dust emission. This supports the interpretation of the ultraviolet emission from the UMa Arc as the product of a process different than dust scattering in the diffuse ISM.   

\end{document}